\documentstyle[11pt,newpasp,twoside,epsf,epsfig,hyperref]{article}
\hypersetup{pdftitle={Methane and the Spectra of T Dwarfs},
  pdfsubject={Methane Dwarfs},
  pdfauthor={Derek Homeier, UGA <derek@physast.uga.edu>}, 
  pdfkeywords={brown dwarfs, stellar atmospheres, infrared spectroscopy}
}
\def\edcomment#1{\iffalse\marginpar{\raggedright\sl#1\/}\else\relax\fi} 
\reversemarginpar 
 
\begin{document} 
\title{Methane and the Spectra of T Dwarfs} 
 \author{Derek Homeier} 
\author{Peter H.~Hauschildt} 
 \affil{Department of Physics \& Astronomy and Center for Simulational 
  Physics, University of Georgia, Athens, GA 30602-2451}
\author{France Allard} 
\affil{CRAL, Ecole Normale 
 Sup{\'e}rieure, Lyon, France} 
 
\begin{abstract} 
We have updated our PHOENIX model atmospheres and theoretical spectra
for ultracool dwarfs with new opacity data for methane based on line
strength predictions with the STDS software. 
By extending the line list to rotational levels of $J\,=\,40$ we can
significantly improve the shape of the near-IR absorption features of
CH$_4$, and in addition find an enhanced blanketing effect, resulting
in up to 50\% more flux emerging in the $J$ band than seen in previous
models, which may thus contribute to the brightening in $J$ and blue
IR colors observed in T dwarfs. 
\end{abstract} 
 
\section{Introduction} 
Absorption bands of H$_2$O and CH$_4$ dominate the IR spectra
of dwarfs cooler than $\sim$\,1400\,K 
and are a backbone of the T dwarf classification schemes defined by
Geballe et al.\ (2002) and Burgasser et al.\ (2002). 
Accurate and complete opacity data for these molecules are thus 
essential for modelling the atmospheres of T dwarfs. 
We have updated the PHOENIX model atmosphere code with a new methane
list based on theoretical calculations 
with the {Spherical Top Data System (STDS)} ({Wenger \& Champion
1998}). Compared to our AMES-Cond models ({Allard et al.\ 2001}),
with $\sim$\,37\,000 CH$_4$ lines from the HITRAN and
GEISA databases, this has added 1.2$\cdot\,10^7$ lines
(Homeier et al.\ 2002). 

\section{Theoretical Spectra}
We have computed opacity sampling spectra in the limiting case of
fully settled dust clouds as described in Allard et al.\ (2001). 
The main difference in the line lists is a much larger 
set of faint lines, mostly due to a coverage in rotational quantum
number extending up to $J$\,=\,40, as compared to typically
$J\,\le\,$10 in the HITRAN and GEISA data. 
We discuss the changes in these models on the case of 
the T\,4.5 dwarf SDSS\,0207+00 observed by Geballe et al.\ (2002). 
As can be seen in the absorption
feature around 2.2\,$\mu$m, rotational levels up to about $J$\,=\,30
need to be included to reproduce the band shape over its full width. 

Less evident in the spectra is the dense background of faint
lines at higher $J$, which effectively closes any windows of small
opacity throughout all of the $H$ and $K$ bands. 
This results in a stronger redistribution of flux towards shorter
wavelengths, and thus a brightening of the $J$ band of about 0\fm5
compared to an AMES-Cond model with the same parameters. 
We note that the colors can be ``fixed'' with lower log\,$g$
models in the AMES-Cond grid, but at the cost of an even worse
spectral fit in the $H$ and $K$ bands. 

\begin{figure}
\plottwo{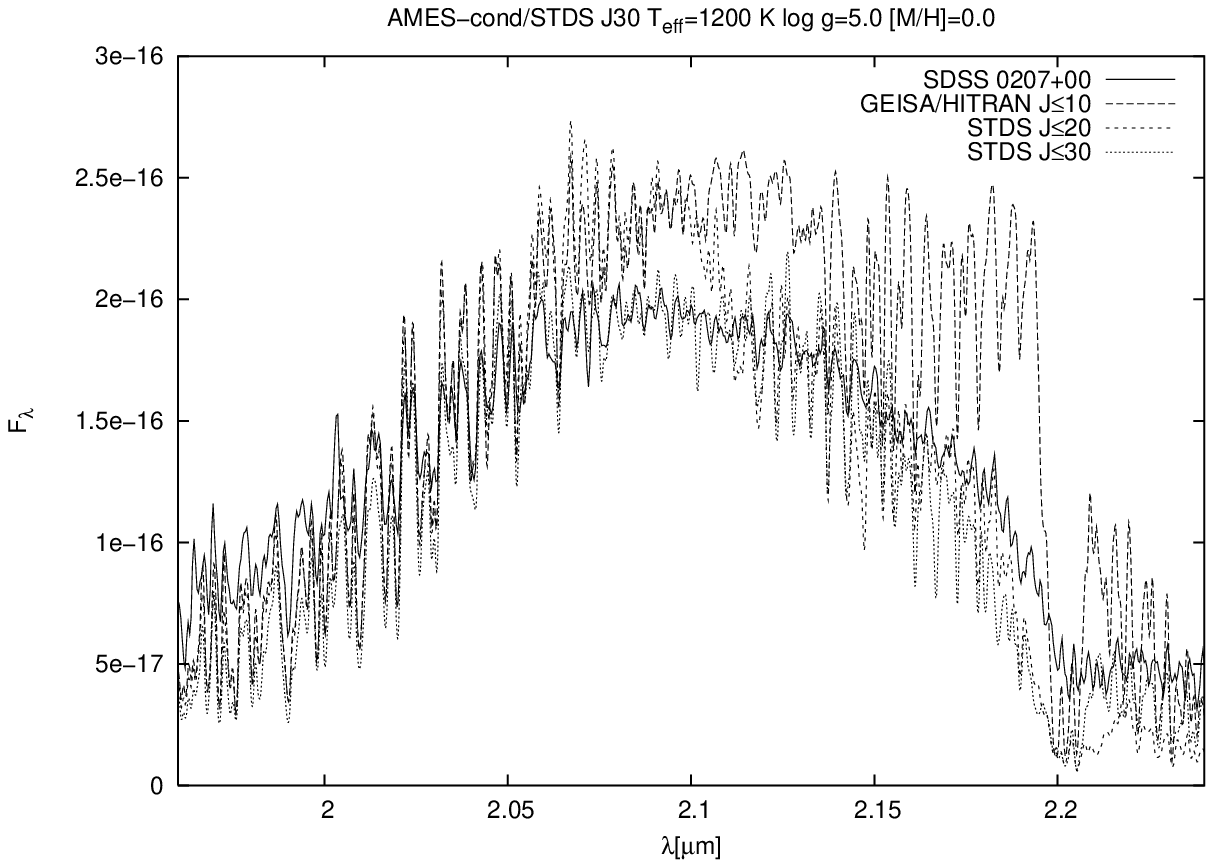}{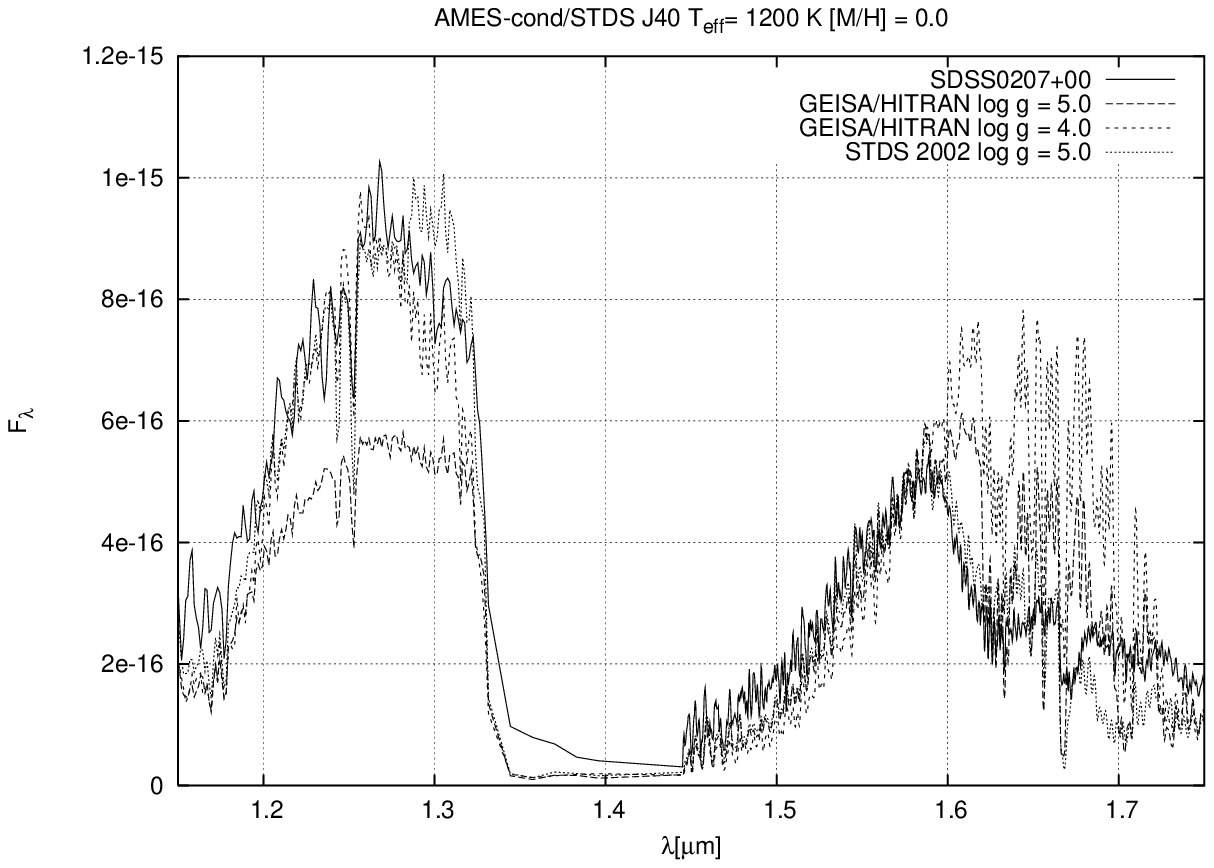}
\caption{AMES-Cond models and new STDS models compared to an observed
T dwarf spectrum: 
$K$ band region with STDS lines up to $J$ of 20 and 30, respectively
(left); $J$ and $H$ band region with log\,$g$\,=\,5.0
models and a log\,$g$\,=\,4.0 AMES-Cond model for comparison
(right).}\vspace{-1.1ex}
\end{figure}

\section{Conclusions}
We could significantly improve our synthetic IR spectra for T dwarfs
with the new STDS line list for methane. 
In addition, the new models produce a shift of emergent flux
towards the $J$ band due to stronger blanketing, reproducing well the
blue IR colors of T dwarfs. To clarify the exact role of 
the onset of CH$_4$ in the rapid brightening in $J$ observed at the
L-T transition, a new set of models will be necessary that also 
provides a consistent treatment of the settling of the cloud layer
below the photosphere expected to occur in about the same temperature
range (Allard et al., these proceedings). 

\vspace*{2.4ex}
\noindent
\textbf{Acknowledgments.} We thank D.\ Alexander, V.\
Boudon, J.-P.\ Champion and U.\ J{\o}rgensen for helpful discussions
and S.\ Leggett for access to observational data. 
This work is supported by NFS grant N-Stars RR185-258.

\end{document}